\documentclass[twocolumn]{aastex631}

\DeclareUnicodeCharacter{2212}{-}
\usepackage{multirow}
\usepackage{comment}
\usepackage[normalem]{ulem}
\usepackage{amsmath}

\shorttitle{Comoving Pairs and the Li-Dip}
\shortauthors{Sun et al.}

\graphicspath{{./}{figures/}}
\begin{document}
		
\title{C3PO V: Comoving stellar pairs indicate rotational spin-down drives the main-sequence Li-Dip}
		
\correspondingauthor{Qinghui Sun}
\email{qinghuisun@sjtu.edu.cn}
		
\author[0000-0003-3281-6461]{Qinghui Sun}
\affiliation{Tsung-Dao Lee Institute, Shanghai Jiao Tong University, Shanghai, 200240, China}
		
\author[0000-0001-5082-9536]{Yuan-Sen Ting}
\affiliation{Department of Astronomy, The Ohio State University, Columbus, OH 43210, USA}
\affiliation{Center for Cosmology and AstroParticle Physics (CCAPP), The Ohio State University, Columbus, OH 43210, USA}
		
\author[0000-0001-8841-3579]{Barbara J. Anthony-Twarog}
\author[0000-0001-5436-5206]{Bruce A. Twarog}
\affiliation{Department of Physics and Astronomy, University of Kansas, 1251 Wescoe Hall Dr., Lawrence, KS, 66045, USA}
		
\author[0000-0003-4794-6074]{Fan Liu}
\affiliation{School of Physics and Astronomy, Monash University, Melbourne, VIC, 3800, Australia}
\affiliation{ARC Centre of Excellence for Astrophysics in Three Dimensions (ASTRO-3D), Canberra, ACT, 2611, Australia}
		
\author[0000-0003-4769-3273]{Yuxi(Lucy) Lu}
\affiliation{Department of Astronomy, The Ohio State University, Columbus, OH 43210, USA}
\affiliation{Center for Cosmology and AstroParticle Physics (CCAPP), The Ohio State University, Columbus, OH 43210, USA}

\begin{abstract}
			
The lithium-dip observed in mid-F dwarfs remains a long-standing challenge for stellar evolution models. We present high-precision stellar parameters and A(Li) for 22 new comoving pairs, primarily located on the hotter side of the Li-Dip. Combined with pairs from the C3PO catalog, our sample includes 124 stars with $T_{\rm eff}$ between 6000 and 7300 K, encompassing and extending slightly beyond the Li-Dip. Among them, 49 comoving pairs (98 stars) have both components within the temperature range of interest. Using this expanded set of comoving pairs observed with high-resolution spectroscopy, we show that rotational spin-down is the dominant process responsible for Li depletion in the Li-Dip. First, within comoving pairs, the star with {\it v} sin {\it i} $>$ 12 km s$^{-1}$ shows higher A(Li) than its more slowly rotating companion within the Li-Dip, indicating that rotation-dependent mixing drives lithium depletion. Second, we observe a correlation between A(Li) and {\it v} sin {\it i}: fast rotators retain higher A(Li) with less scatter, while slow rotators show lower A(Li) and greater dispersion. Third, among slow rotators, A(Li) varies widely, suggesting that differences in initial rotation rates and spin-down histories influence how much Li is depleted. Some stars may have formed as fast rotators and spun down rapidly, leading to more Li depletion, while others may have started as slow rotators and retained more of their initial Li. These results demonstrate that rotational induced mixing plays an important role in shaping the Li-Dip beyond the effects of stellar age and mass.
			
\end{abstract}
		
\section{Introduction}
		
Lithium (Li), along with Beryllium (Be) and Boron (B), exists only in the outermost layers of main-sequence (MS) stars. These elements are fragile, as they are destroyed at relatively low temperatures compared to heavier elements. Consequently, their surface abundances serve as valuable tracers of the internal processes occurring below a star's surface. The standard stellar evolution theory (SSET; e.g., \citealt{1967ARA&A...5..571I, 1990ApJS...73...21D, 2017AJ....153..128C}) provides a basic framework for understanding Li behavior, but it assumes stars evolve without the influence of complex factors such as rotation, magnetic fields, diffusion, and mass loss.
		
According to the SSET, the extent of Li depletion should primarily depend on the depth of a star's convective envelope. Stars like the Sun (G-type dwarfs) and cooler K-type dwarfs are predicted to lose some of their surface Li early in their evolution, as their deep convective envelopes transport Li into hot interior regions where it is rapidly destroyed by nuclear reactions. This process is expected to occur mostly during the pre-main sequence phase. In contrast, hotter F- and A-type stars have much shallower convective zones, so the SSET predicts that their surface Li should remain largely unaltered throughout their MS lifetimes.
		
Observations, however, reveal a more complex picture. A striking anomaly has been discovered among mid-F stars, where a sharp decline in surface Li abundance, known as the ``Li-Dip", has been observed (\citealt{1986ApJ...302L..49B}). This unexpected depletion could not be explained by standard convective mixing, suggesting that important physical processes have been missing from the SSET. Subsequent studies have found further intriguing properties: not only do mid-F stars experience notable Li depletion, but cooler G and K dwarfs also show more Li depletion over their MS lifetimes (\citealt{1995MNRAS.273..559J, 2005MNRAS.358...13J}). More recently, evidence has emerged that even A-type dwarfs show Li depletion over time (\citealt{2019AJ....158..163D}). These findings suggest that Li over-depletion in MS stars is not limited to a specific spectral type but is a widespread phenomenon affecting even stars of a higher mass where surface convection on the main sequence is weak to nonexistent.
		
Several mechanisms have been proposed to explain these discrepancies, including diffusion (\citealt{1993ApJ...416..312R}), mass loss (\citealt{1992ApJ...395..654S}), and slow internal mixing driven by gravity waves (\citealt{2005Sci...309.2189C}) or rotation (\citealt{1989ApJ...338..424P}). Among these, rotationally induced mixing has emerged as a promising explanation. Some observations show that Li depletion is closely tied to stellar spin-down: as stars lose angular momentum over time, internal mixing processes redistribute material between the surface and deeper layers, where Li is gradually destroyed. This connection is evident in the correlated depletion of Li with Be and B (\citealt{1998ApJ...498L.147D, 2004ApJ...613.1202B, 2016ApJ...830...49B}), as well as with the slowing rotation rates of stars (\citealt{2019AJ....158..163D}). The specific type of rotational mixing supported by these observations involves angular momentum loss triggering shear instabilities, which in turn drive internal mixing and lead to progressive Li depletion (\citealt{1989ApJ...338..424P, 1997ApJ...488..836D, 2015MNRAS.449.4131S}).
		
Extensive studies of open clusters have explored the role of rotational spin-down in shaping the Li-Dip. Observations of clusters including the Pleiades (120 Myr), M48 (420 Myr), and Hyades/Praesepe (650 Myr) show progressive Li depletion with age in the effective temperature range of 6735 to 6200 K, where the Li-Dip is most pronounced \citep{2023ApJ...952...71S}. Several studies have supported rotational mixing as the main mechanism responsible for this depletion during main-sequence evolution \citep[e.g.,][]{2012AJ....144..137C, 2017AJ....153..128C}. As stars lose angular momentum through their outer layers, they experience differential rotation and spin-down, leading to gradual Li depletion correlated with age and rotation. More recently, \citet{2025arXiv250704266S} found that rotationally induced mixing can create the main-sequence Li-Dip by studying subgiants in NGC 188. Despite this progress, important questions remain about the role of stellar spin-down and rotational history in driving Li depletion. Addressing these uncertainties needs further observational constraints.
		
Comoving pairs of stars which share a common origin and similar stellar properties, may exhibit different internal structures and/or rotational histories, thereby providing a powerful complementary approach to open cluster studies. In this study, we leverage a sample of 124 stars covering the Li-Dip and extending slightly beyond (7300 - 6000 K), including 49 comoving pairs (98 stars), to investigate the mechanisms driving the Li-Dip and shed light on the unresolved questions surrounding Li depletion in stars.
		
\section{Data Selection} 
		
We first incorporate high-precision A(Li) measurements from C3PO (Complete Census of Co-moving Pairs of Stars) studies \citep{2023MNRAS.526.2181Y, 2024Natur.627..501L, 2025ApJ...978..107S}, which include 125 comoving pairs of stars (250 stars total). The comoving pairs are selected from the Gaia EDR3 catalog \citep{2021A&A...649A...1G} based on small spatial separations ($\lesssim$ 10$^7$ AU), and are therefore presumed to have formed in the same environment. While a more conservative threshold of $10^6$ AU has been suggested \citep{2023MNRAS.526.2181Y}, we adopt $10^7$ AU here to enlarge the sample. We note, however, that this choice may introduce some chance-alignment contaminants, and future follow-up observations are encouraged.
		
These spectra have been obtained using the Magellan, Keck, and Very Large Telescopes, with resolutions of R $\sim$ 50,000–110,000 and typical S/N $\sim$ 250 per pixel. For this study, we select 87 stars from the C3PO sample with $T_{\rm eff}$ $>$ 6000 K. Among these, we find 32 co-moving pairs (64 stars), while the remaining 23 stars have companions that do not meet $T_{\rm eff}$ $>$ 6000 K.
		
\begin{deluxetable*}{lllcccccccccc}
			\label{Table1}
			\tablecaption{Stellar ID, coordinates, and astrometry of the 49 co-moving pairs}
			\tabletypesize{\small}
			\tablewidth{1.0\textwidth}
			\setlength{\tabcolsep}{0.8pt}
			\tablehead{
				\colhead{Name$^a$} & \colhead{ID$^a$} & \colhead{Simbad ID$^a$} & \colhead{RA$^a$} & \colhead{DEC$^a$} & \colhead{G$^b$} & \colhead{Bp-Rp$^b$} & \colhead{pmRA$^b$} & \colhead{epmRA$^b$} & \colhead{pmDEC$^b$} & \colhead{epmDEC$^b$} & \colhead{...$^b$} & \colhead{pmem$^c$} \\
				\colhead{} & \colhead{} & \colhead{} & \colhead{(deg)} & \colhead{(deg)} & \colhead{(mag)} & \colhead{(mag)} & \colhead{(mas yr$^{-1}$)} & \colhead{(mas yr$^{-1}$)} & \colhead{(mas yr$^{-1}$)} & \colhead{(mas yr$^{-1}$)} & \colhead{...} & \colhead{} 
			}
			\startdata
			sk183a	& 692119656035933568 & HD 126766 & 137.099 & 27.535 & 8.121 & 0.732 & -53.218 & 0.025 & 71.761 & 0.019 & ... & field \\
			sk184b	& 692120029700390912 & BD+28 1698 & 137.113 & 27.543 & 8.144 & 0.738 & -51.770 & 0.023 & 73.733 & 0.017 & ... & field \\
			\hline
			sk177a	& 704994524881597184 & BD+29 1836B & 133.186 & 28.458 & 9.851 & 0.734 & -46.744 & 0.018 & 29.103 & 0.015 & ... & field \\
			sk178b	& 704994524881597056 & BD+29 1836A & 133.188 & 28.458 & 9.855 & 0.733 & -45.666 & 0.016 & 29.366 & 0.014 & ... & field \\
			... & ... & ... \\
			\hline
			\enddata
			\tablecomments{a. Our designated name, Gaia DR3 ID, and simbad name of the star, RA and DEC in degrees (J2000). \\
				b. Gaia $G$ magnitude, $B_p-R_p$ color, proper motion in RA, DEC, and errors from Gaia, radial velocity, parallax, and errors from Gaia. The spatial separation of the pair in AU ($\Delta\ s$). \\
				c. Membership probability of 27 nearest young associations within 150 pc of the Sun, calculated by using the BANYAN $\Sigma$ code (\citealt{2018ApJ...856...23G}), using the location and astrometry of the star. Some of the stars have high probability of belonging to the Argus (ARG), Lower Centaurus Crux (LCC), Upper Centaurus Lupus (UCL), Upper Scorpius region (USCO),$\beta$ Pictoris Moving Group (BPMG) associations. \\
				(This table is available in its entirety in machine-readable form online.)
}
\end{deluxetable*}
		
We have also further obtained new high-resolution spectroscopic data for 22 comoving pairs (44 stars) using the Magellan/MIKE II spectrograph \citep{2003SPIE.4841.1694B} during observing runs in August and November 2021. These targets are selected from the Gaia EDR3 catalog in a manner similar to the C3PO sample. The selection focuses on stars on the hotter side of the Li-Dip and allows for larger spatial and radial velocity separations, given the relative small number of comoving systems among hotter stars. Details of the 22 comoving pairs, including their designated IDs, SIMBAD names, right ascension, and declination (J2000), are provided in Table~\ref{TableA1}. We select 37 stars with effective temperatures $T_{\rm eff} > 6000$ K from the new sample. Among these, 17 pairs (34 stars) are confirmed comoving pairs. One star has a cooler companion ($T_{\rm eff}$) and is therefore excluded, while star4-b and star29-a are separated by too large a distance to be considered comoving pairs. These newly observed pairs primarily span the hotter side of the Li-Dip and several A-type dwarfs.
		
For the new data shown in this work, observations have been conducted using the 0.5" slit, achieving spectral resolutions of approximately 44,000 in the red and 57,000 in the blue. The wavelength coverage spans 3350–5000 \AA\ in the blue and 4900–9500 \AA\ in the red, with all spectra achieving a signal-to-noise ratio (S/N) greater than 100 near the Li I 6707.8 \AA\ line. Calibration frames, including quartz flats, milky flats, and fringe flats, have been obtained at the beginning of each night, while ThAr lamp spectra have been regularly acquired before and after science exposures. All MIKE spectra are processed using the CarPy standard pipeline\footnote{https://github.com/alexji/mikerun} \citep{2000ApJ...531..159K, 2003PASP..115..688K}, which performs key reduction steps including overscan correction, flat-fielding, wavelength calibration, and the combination of spectra into final red- and blue-side spectra. We further normalize the continuum of the combined spectra to ensure consistency across all stars. Both C3PO and the newly added data in the study have undergone the same spectroscopy and Li abundance analysis, ensuring consistency.
		
Combining both sources, our final sample consists of 124 stars: 87 from the C3PO dataset and 37 from our new MIKE observations (note that only those with $T_{\rm eff}$ $>$ 6000 K is included). This sample includes 49 comoving pairs (98 stars), while the remaining stars have companions that are either cooler than 6000 K or located too far apart to be considered comoving (one pair). This carefully selected sample enables us to investigate A(Li) patterns in comoving pairs that share the same origin, similar parameters, with $T_{\rm eff}$ spanning the whole Li-Dip range and extending slightly beyond (6000 - 7000 K). Table \ref{Table1} lists the star IDs, Gaia $G$ magnitudes, $B_p - R_p$ colors, proper motions in RA and DEC, and other astrometric parameters from Gaia for the 50 comoving pairs. Using these coordinates and astrometric data, we compute the probability that each pair belongs to a young stellar association with the BANYAN $\Sigma$ package \citep{2018ApJ...856...23G}. The membership probability is provided in the last column of Table \ref{Table1}.
		
\section{Stellar Parameters}
		
Table \ref{Table2} presents the stellar atmospheric parameters, projected rotational velocities ({\it v} sin {\it i}), ages, and lithium abundances (A(Li)) for the 49 comoving pairs analyzed in this study. Of these, 22 new comoving pairs (44 stars) are listed in Table \ref{TableA1}, including several cooler stars not discussed further in this paper. The parameters for these new pairs were derived using the methods described in \citet{2025ApJ...978..107S, 2025ApJ...980..179S}, which are briefly summarized in the subsections below. For the remaining pairs drawn from the C3PO sample, we adopt published values from \citet{2023MNRAS.526.2181Y, 2024Natur.627..501L, 2025ApJ...978..107S}.

\begin{deluxetable*}{lcccccccccccccc}
	\label{Table2}
	\tablecaption{Stellar Parameters and A(Li) of the 49 co-moving pairs}
	\tabletypesize{\small}
	\tablewidth{0.95\textwidth}
	\setlength{\tabcolsep}{1.5pt}
	\tablehead{
		\colhead{Name$^a$} & \colhead{$ID^a$} & \colhead{Simbad ID$^a$} & \colhead{$T_{\rm eff}^b$} & \colhead{log $g^b$} & \colhead{[Fe/H]$^b$} & \colhead{$V_t^b$} & \colhead{$v$ sin $i ^c$} & \colhead{e$v$ sin $i ^c$} & \colhead{$P_{\rm ROT}^c$} & \colhead{Age$^c$} & \colhead{$\sigma_{\rm Age}^c$} & \colhead{S/N$^c$} & \colhead{A(Li)$^c$} & \colhead{...} \\
		\colhead{} & \colhead{} & \colhead{} & \colhead{(K)} & \colhead{(dex)} & \colhead{(dex)} & \colhead{(km s$^{-1}$)} & \colhead{(km s$^{-1}$)} & \colhead{(km s$^{-1}$)} &\colhead{(days)} & \colhead{(Gyr)} & \colhead{(Gyr)} & \colhead{} & \colhead{(dex)} & \colhead{...}
	}
	\startdata
	sk183a	& 692119656035933568 & HD 126766 & 6003 & 4.56 & -0.355 & 1.04 & 20.5 & 1.17 & -- & 0.6 & 0.5 & 331 & 2.62 & ... \\
	sk184b	& 692120029700390912 & BD+28 1698 & 6004 & 4.60 & -0.333 & 1.06 & 20.39 & 1.16 & -- & 0.39 & 0.32 & 310 & 2.62 & ... \\
	\hline
	sk177a	& 704994524881597184 & BD+29 1836B & 6034 & 4.49 & -0.289 & 1.13 & 20.52 & 1.26 & -- & 1.37 & 0.7 & 315 & 2.48 & ... \\
	sk178b	& 704994524881597056 & BD+29 1836A & 6070 & 4.53 & -0.291 & 1.18 & 20.58 & 1.27 & -- & 0.7 & 0.46 & 323 & 2.49 & ... \\
	\hline
	... & ... & ... \\
	\enddata
	\tablecomments{a. Our designated name, Gaia DR3 ID, and simbad name of the pair. The last column indicates whether the stellar atmosphere parameters are derived using FASMA spectral synthesis or from standard ionization balance. \\
		b. Stellar atmosphere parameters, including $T_{\rm eff}$, log {\it g}, $V_t$, and their associated uncertainties. \\
		c. Rotational velocity and associated uncertainty of the star; rotational period derived from stellar light curve and from \citet{2024AJ....167..189C}; age and associated uncertainty derived from spectroscopic log g. Signal-to-noise ratio (S/N) near the Li I 6707.8 \AA doublet. The 1D-LTE A(Li), if the number starts with a ``$<$", it means an upper limit A(Li), else it is a detection. If an detection, the uncertainty associated with A(Li) is given. \\
		(This table is available in its entirety in machine-readable form online.)
	}
\end{deluxetable*}
		
\subsection{Stellar Atmosphere}
		
The stellar atmospheric parameters include effective temperature ($T_{\rm eff}$), surface gravity (log g), microturbulence ($V_t$), and metallicity ([Fe/H]). For slow-rotating stars ({\it v} sin {\it i} $<$ 30 km s$^{-1}$), we follow the methods described in \citet{2025ApJ...980..179S} to determine these parameters and uncertainties. We measure equivalent widths (EWs) for Fe I and Fe II lines using the \textit{splot} task in IRAF\footnote[1]{IRAF is distributed by the National Optical Astronomy Observatories, operated by the Association of Universities for Research in Astronomy Inc., under a cooperative agreement with the National Science Foundation.}, and then use the q$^2$ Python package \citep{Ramirez2014} to derive the stellar atmospheric parameters, using the MARCS model atmosphere grid \citep{2008A&A...486..951G}. Excitation equilibrium of Fe I lines is enforced to determine $T_{\rm eff}$, while ionization balance between Fe I and Fe II lines is used to constrain log g. Microturbulence ($V_t$) is derived by minimizing trends between iron abundance and reduced EWs, using an iterative approach to ensure parameter convergence. 
		
For rapidly rotating stars ($v \sin i > 30$ km s$^{-1}$), spectral line blending becomes significant, making spectrum synthesis more reliable than ionization balance methods for deriving stellar atmospheric parameters. For such stars, we use the FASMA synthesis package \citep{2018MNRAS.473.5066T}, which wraps the MOOG spectral synthesis code and adopts the MARCS model atmosphere grid. In each pair where at least one component has $v \sin i > 30$ km s$^{-1}$, we apply FASMA synthesis to both stars to avoid systematic differences introduced by using different methods within a pair. The last column of Table \ref{Table2} indicates whether FASMA synthesis is used. As expected, hotter stars with higher $v \sin i$ tend to show larger parameter uncertainties than cooler, slowly rotating stars.
		
\subsection{Projected Rotational Velocity and Rotational Period}
		
We compute projected rotational velocity ({\it v} sin {\it i}) by using the \textit{fxcor} task in IRAF, following the methods of \citet{2020AJ....159..220S, 2022MNRAS.513.5387S, 2025ApJ...978..107S}. This method involves performing a cross-correlation between the observed spectrum and a template in Fourier space, fitting a Gaussian to the highest peak of the cross-correlation function, and determining line broadening based on the peak's width and central position \citep{1984A&A...138..183B}. We derive {\it v} sin {\it i} and its uncertainty from the spectral line broadening.
		
We further derive rotational periods ($P_{\rm ROT}$) from publicly available TESS light curves \citep{Stassun2018_TIC1, Stassun2019_TIC2}, using the \texttt{unpopular} package \citep{2022AJ....163..284H}. To account for potential blending from nearby sources, we vet each period using \texttt{TESS\_localize} \citep{2023AJ....165..141H}. We also cross-match our sample with several literature catalogs. No matches were found in the ZTF periodicity catalog\footnote{\url{http://atua.caltech.edu/ZTF/Zubercal.html}} \citep{2022AJ....164..251L, 2024AJ....167..159L}, or in the Kepler rotation period catalogs \citep{2021ApJS..255...17S, 2014ApJS..211...24M}. Several stars overlap with two TESS-based $P_{\rm ROT}$ catalogs \citep{2022ApJ...936..138H, 2024AJ....167..189C}; where available, their reported values are in agreement with our derived $P_{\rm ROT}$ within 2$\sigma$. Table~\ref{Table2} reports the $P_{\rm ROT}$ values for the 50 comoving pairs when they can be reliably derived from TESS light curves or found in the literature. However, the majority of stars in our sample still lack $P_{\rm ROT}$ measurements, and thus $P_{\rm ROT}$ are not used for further analysis, as the completeness is too low to support any statistical conclusions.
		
\subsection{Stellar Age}
		
We identify several pairs as members of young associations using BANYAN~$\Sigma$, as shown in Table~\ref{Table1}. These stars are likely very young pre-MS (PMS) objects, with ages ranging from a few Myr to several tens of Myr. For these systems, we adopt the age of the associated young stellar group and list it in Table~\ref{Table2}. These stars all show high A(Li), with some reaching or matching the meteoritic value of 3.3 dex.
		
Stellar ages are determined using the q$^2$ Python routine \citep{Ramirez2014} and the Yonsei-Yale (Y$^2$) isochrones \citep{2003ApJS..144..259Y}. We use spectroscopically determined log g to place each star on the Hertzsprung-Russell diagram and compare it with theoretical stellar evolution models. The precision of this method depends on the accuracy of log g, which is influenced by uncertainties in Fe I and Fe II line measurements. Age determinations near solar metallicity remain uncertain due to empirical isochrone variations \citep[e.g.][]{1993ApJS...86..153G}. A maximum-likelihood estimation is applied to determine the most probable stellar age, with values and uncertainties provided in Table~\ref{Table2} and \ref{TableA1}.
		
\begin{figure}
			\centering	\includegraphics[width=0.45\textwidth]{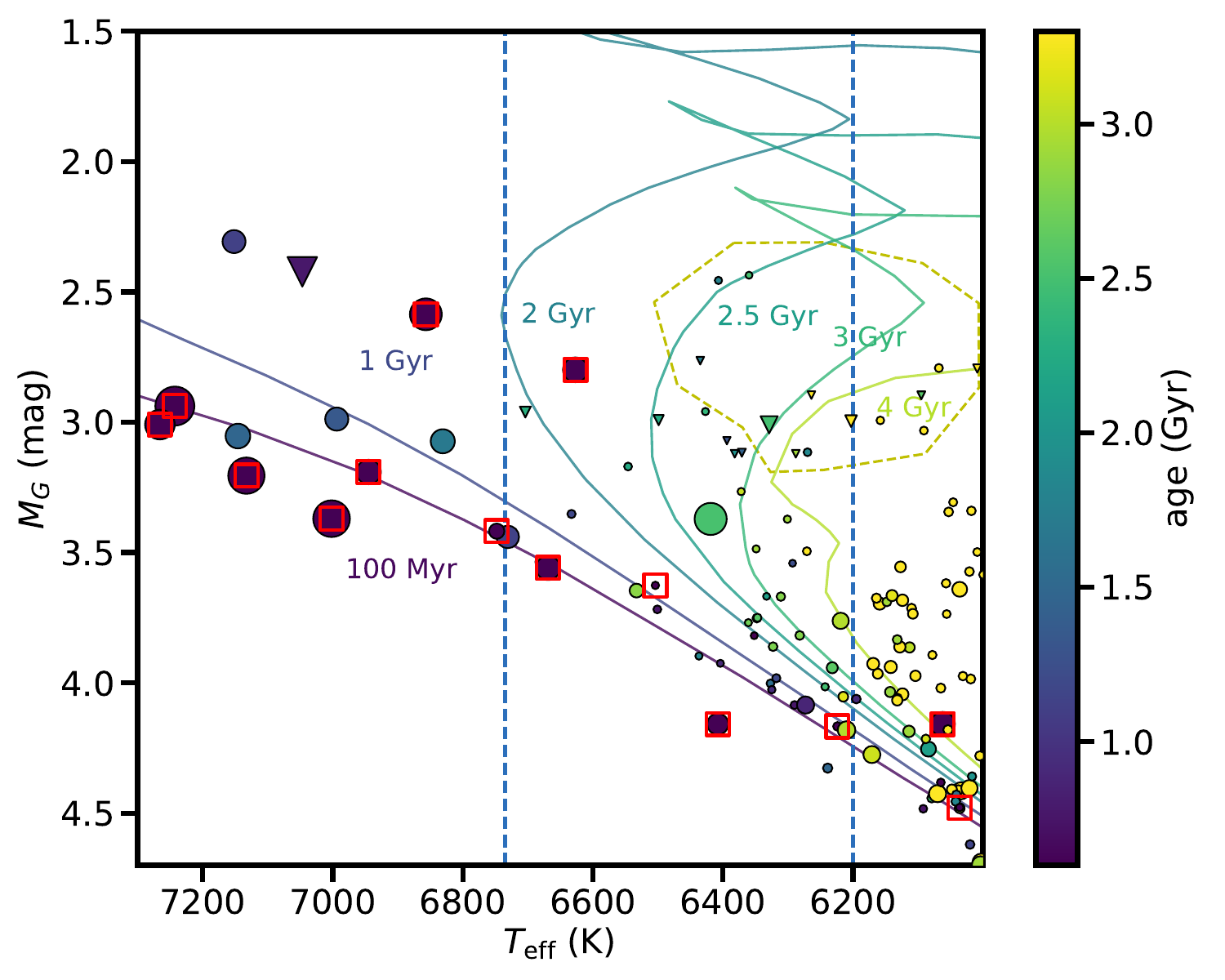}
			\caption{The H-R diagram for comoving pairs, the Li-Dip $T_{\rm eff}$ range (6735-6200 K) is denoted. The absolute $G$ magnitudes are computed from Gaia apparent $G$ magnitudes and parallaxes (\citealt{2018ipac.data..I12C}). Comoving pairs are color-coded by age, with PARSEC isochrones (\citealt{2022A&A...665A.126N}) using Gaia passbands from \cite{2018A&A...616A...4E} for stellar tracks at 100 Myr, 1 Gyr, 2 Gyr, 2.5 Gyr, 3 Gyr, and 4 Gyr. The symbol sizes are proportional to {\it v} sin {\it i}. The filled circles are Li detections, downward triangles are 3$\sigma$ upper limit A(Li). PMS stars from young associations are marked as red squares, and evolved stars are enclosed in yellow dashed-line circles; both are excluded from further Li-Dip analysis.}
			\label{Fig1}
\end{figure}

With the stellar parameters derived for the comoving pairs, Figure~\ref{Fig1} shows an H-R diagram of the 124 stars with $T_{\rm eff}$ $>$ 6000 K. Since the Li-Dip is defined only for MS stars, and that evolved stars can artificially ``recreate" the MS Li-Dip, we identify and mark likely contaminants: PMS stars belonging to young associations are shown as red squares, and more evolved stars are enclosed in yellow dashed-line circles. These contaminating sources are excluded from further discussion of the Li-Dip. Note that this HR diagram is used solely to select MS stars, and log g are derived from spectroscopy rather than photometry.

\subsection{Lithium Abundance}
		
A detailed description of the derivation of A(Li) is available in \citet{2022MNRAS.513.5387S, 2023ApJ...952...71S, 2025ApJ...978..107S} and is summarized here. The 1D-LTE A(Li) are determined by synthesizing the Li I 6707.8 \AA\ doublet using the \textit{synth} task in MOOG. The line list near the Li feature is adopted from \citet{2022MNRAS.513.5387S}, which includes the nearby Fe I line and relevant molecular lines, calibrated using cool stars as described in that study. Errors are propagated in quadrature from uncertainties in equivalent width measurements and stellar atmospheric parameters. For stars rotating faster than 30 km s$^{-1}$, a rotationally broadened profile is used instead of a Gaussian core for spectral synthesis.
		
For weak Li lines, we distinguish between detections and upper limits. The 3$\sigma$ EW threshold is computed using signal-to-noise ratio (S/N), full width at half maximum, and instrument pixel scale, following the method of \citet{1993ApJ...414..740D}. Li detections are required to have a Li line stronger than the A(Li) corresponding to the 3$\sigma$ EW threshold. For detections, non-local thermal equilibrium (NLTE) corrections are computed using the BREIDABLIK package, as shown in Table \ref{TableA1}. Throughout this paper, we adopt 1D-LTE A(Li) for discussing the Li-Dip, as the 3$\sigma$ upper limit A(Li) values are essential for discussing Li depletion yet not corrected for NLTE effect. Also, when investigating differential Li depletion in the Li-Dip, NLTE corrections are expected to be nearly uniform across stars within a narrow parameter range investigated in this study, making their impact to be minimal for the results. The final adopted 1D-LTE A(Li) for the 49 comoving pairs are shown in Table \ref{Table2}.
		
\section{Results}
		
\begin{figure*}
			\centering	\includegraphics[width=1.0\textwidth]{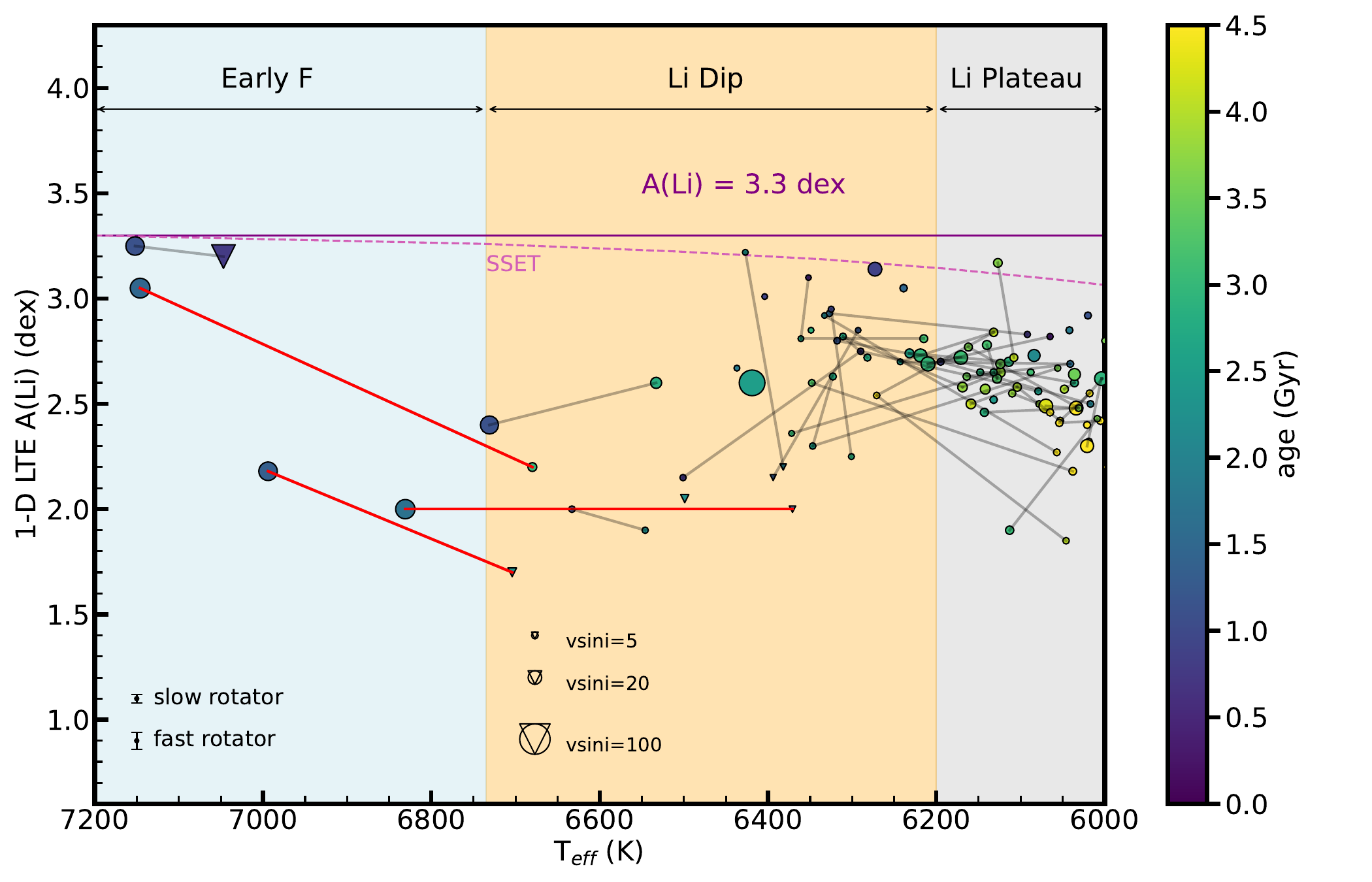}
			\caption{The distribution of the MS Li-Dip and slightly beyond. The Li-Dip spans approximately 6735 K to 6200 K, characterized by significant Li deficiency relative to the adjacent early-F and Li-Plateau regions. The meteoritic A(Li) = 3.3 dex is marked in the plot with a purple line as reference. The selected sample are shown as filled circles, color-coded by age. The comoving pairs are connected by gray lines, while the three pairs consisting of one star hotter than 6735 K and one cooler are highlighted with red lines. Symbol sizes are proportional to {\it v} sin {\it i}. The SSET model for A(Li) depletion as a function of $T_{\rm eff}$ is shown as a pink dashed line.}
			\label{Fig2}
\end{figure*}

Figure~\ref{Fig2} presents the A(Li)-$T_{\rm eff}$ distribution for comoving pairs hotter than 6000 K. Each comoving pair is connected by a gray line. Each of the comoving pairs is connected by a gray line; pairs not linked in the figure contain one companion cooler than 6000 K, or one companion excluded as evolved stars. Pairs that are likely members of a young stellar association are excluded. Following the definitions in \citet{2023ApJ...952...71S} and several open cluster studies that clearly delineate the Li-Dip \citep{2009AJ....138..159H, 2019AJ....158..163D}, we define early F-type dwarfs as stars with $T_{\rm eff}$ between 7200 K and 6735 K, located on the hot side of the Li-Dip. The Li-Dip itself spans 6735 K to 6200 K and is characterized by a pronounced depletion in A(Li) relative to nearby $T_{\rm eff}$ regions. A short Li plateau lies between 6200 K and 6000 K, where stars typically show less lithium depletion.

We adopt the meteoritic A(Li) value of 3.3 dex for comparison \citep{1989GeCoA..53..197A}. The primordial lithium abundance from Big Bang nucleosynthesis is approximately 2.6 dex \citep{2003ApJS..148..175S}, and subsequent lithium enrichment through red giant evolution and other Galactic processes \citep[e.g.,][]{2022MNRAS.513.5387S} increases the interstellar A(Li) to 3.3 dex, consistent with the meteoritic value. Given this enrichment in the star-forming environment, we adopt the meteoritic A(Li) as the reference for assessing Li depletion in stars, rather than the primordial value.
		
\subsection{Early F dwarfs (7200 - 6735 K)}
		
In the temperature range slightly hotter than the Li-Dip, all five stars show {\it v} sin {\it i} above 35 km s$^{-1}$, while their A(Li) span a broad range from 2.0 to 3.25 dex. Three of these stars have A(Li) measurements or upper limits above 3.0 dex, with ages of 1.48, 1.13, and 0.77 Gyr and subsolar metallicities. The remaining two stars show A(Li) values slightly above or equal to 2.0 dex, with ages of 1.34 and 1.70 Gyr and slightly super-solar metallicities. Compared to the meteoritic A(Li) of 3.3 dex, all five stars show evidence of Li depletion, with reductions ranging from approximately 0.05 to 1.3 dex. These results suggest that lithium depletion in early-F dwarfs can begin as early as a few hundred million years and is not driven by age alone.
		
\subsection{The MS Li-Dip (6735 - 6200 K)}

The Li-Dip is a well-established, non-standard phenomenon characterized by significant Li depletion in mid-F dwarfs \citep{2009AJ....138..159H, 2019AJ....158..163D, 2025arXiv250704266S}. A(Li) drops sharply at the hot side of the Li-Dip ($\sim$ 6700 K), from approximately 2.5 dex to below 1.7 dex. More rapid rotators also tend to show higher A(Li). Moving from the hot side of the Li-Dip toward cooler $T_{\rm eff}$, the comoving pairs show a wide range in A(Li), from detections around 3.2 dex to upper limits below 2.1 dex. Some stars follow the SSET model closely, while others fall below it. A mixture of stellar ages and {\it v} sin {\it i} values is evident at the cooler side of the Li-Dip, and the A(Li) scatter is notably larger than that seen in the Li plateau. Together, these features characterize the Li-Dip, marked by over-depletion and large spread in A(Li) among mid-F dwarfs in the 6735–6200 K range.

In our sample, we identify three comoving pairs in which one star is hotter than 6735 K, rotates rapidly, and retains relatively high A(Li), while its companion lies within the Li-Dip, rotates slowly, and shows lower A(Li) detections or only upper limits. These pairs are highlighted with red lines in Figure~\ref{Fig2}. These differences suggest that rotation plays an important role in driving Li depletion within the Li-Dip, even among comoving pairs with similar $T_{\rm eff}$ and shared formation histories. Although the stars in each pair may differ slightly in $T_{\rm eff}$, these differences alone cannot account for the observed discrepancies in A(Li). Compared to the predicted depletion trend from the SSET model \citep{2015MNRAS.449.4131S}, the observed Li differences within the paired stars are much larger than expected based on their $T_{\rm eff}$ offsets. The SSET model predicts less than 0.01 dex of A(Li) depletion from 7200 to 6735 K, indicating that rotation is needed to explain the observed behavior. We also note that the comoving pairs presumably share a common origin and thus have the same age, although our isochrone-based age estimates in the earlier section suggest differences. While measurement uncertainties likely contribute to these discrepancies, we note that in the three red-line-linked pairs, the low-A(Li) companion has an age that is similar to the fast-rotating companion.
		
\subsection{The Li-Plateau}
		
The Li-Plateau is between 6200 to 6000 K \citep{2023ApJ...952...71S}, and is characterized by the relative retention of Li compared to the depletion seen in the Li-Dip and in cooler G/K dwarfs. In our sample, we observe a gradual decrease in A(Li) with increasing age. Beyond the upper boundary of the Li-Plateau (around 6200 K), A(Li) shows increased scatter, while within the Li-plateau, A(Li) remains relatively consistent with small variations at a given age.
		
\subsection{Correlation Between A(Li) and {\it v} sin {\it i}}
		
\begin{figure*}
			\centering	\includegraphics[width=1.0\textwidth]{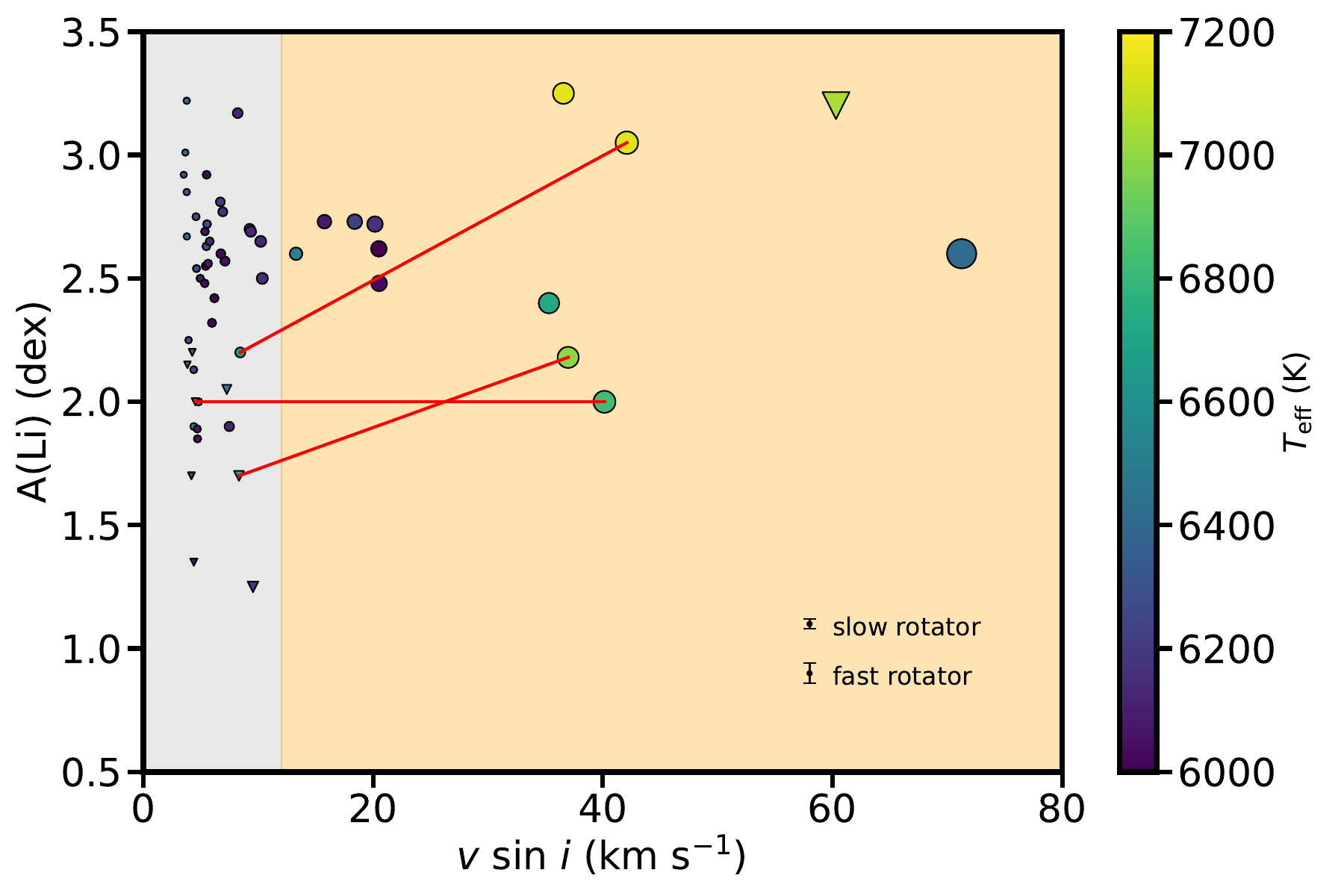}
			\caption{A(Li) as a function of {\it v} sin {\it i}, color-coded by $T_{\rm eff}$. The symbol sizes are scaled proportionally to {\it v} sin {\it i}. Two regions with {\it v} sin {\it i} $<$ 12 km s$^{-1}$ and {\it v} sin {\it i} $>$ 12 km s$^{-1}$ are shaded with different background colors.}
			\label{Fig3}
\end{figure*}

Our comoving pairs are particularly useful for examining the relationship between A(Li) and {\it v} sin {\it i}, as they consist of stars with shared origins and initial compositions but different rotational histories. The differences in A(Li) between companions can therefore be more confidently attributed to rotation, rather than to variations in mass, age, or initial composition. Figure~\ref{Fig3} shows A(Li) as a function of {\it v} sin {\it i}. A(Li) clearly decreases with decreasing {\it v} sin {\it i}, particularly for stars with {\it v} sin {\it i} below 12 km s$^{-1}$. Stars with $v \sin i > 12$ km s$^{-1}$ generally maintain relatively high A(Li) and show smaller scatter compared to those with slower rotation.
		
Below {\it v} sin {\it i} $\sim$ 12 km s$^{-1}$, A(Li) spans a wide range from upper limits below 1.3 dex to 3.2 dex. The large intrinsic scatter among slow rotators likely reflects diverse initial rotation rates and spin-down histories: some stars may have begun as fast rotators, depleting more Li as they spun down, while others were born slow and retained more of their initial Li. In contrast, fast rotators show relatively tight A(Li) distributions, emphasizing that rotational history, rather than age or mass alone, is the key driver of A(Li) and the development of the Li-Dip. Our comoving pairs achieve lower detection limits for both {\it v} sin {\it i} and A(Li) compared to previous open cluster studies, particularly improving measurements in the slower rotating stars.
		
The analysis of comoving pairs offers a direct test of the Li--rotation connection by comparing stars of similar mass and age. In Figure~\ref{Fig3}, pairs with one component above (high A(Li)) and the other below (low A(Li)) {\it v} sin {\it i} = 12 km s$^{-1}$ are linked by gray lines, illustrating how rotational spin-down leads to Li depletion in stars with shared evolutionary histories. The clearest correlations between {\it v} sin {\it i} and A(Li) are seen in stars around $T_{\rm eff} \sim 6700$ K, where Li depletion is most severe. By the age of Hyades/Praesepe ($\sim$650 Myr), most stars in this temperature range have already depleted Li to levels below the detection limit in past open cluster studies. In contrast, our comoving pairs show detectable Li even though they are slightly older, likely due to the lower detection limits achieved with our high-resolution, high-S/N Magellan/MIKE spectra compared to previous open cluster work.
	
At around 6700 K, rotational mixing appears to be the only effective mechanism for Li depletion. Neither convective overshoot nor magnetic braking alone can account for the observed patterns. Stars near 6700 K have radiative envelopes and very shallow surface convection zones. Convective overshoot, which requires a deep convective envelope to transport material into hotter interior layers where Li can be destroyed, is ineffective in such stars. Additionally, magnetic braking operates through magnetized stellar winds, which are generated by the interaction between surface magnetic fields and outer convection zones. Stars around 6700 K lack sufficiently deep convection zones to sustain strong magnetic fields, making magnetic braking inefficient at producing the observed Li depletion.
	
\begin{figure*}
	\centering	\includegraphics[width=1.0\textwidth]{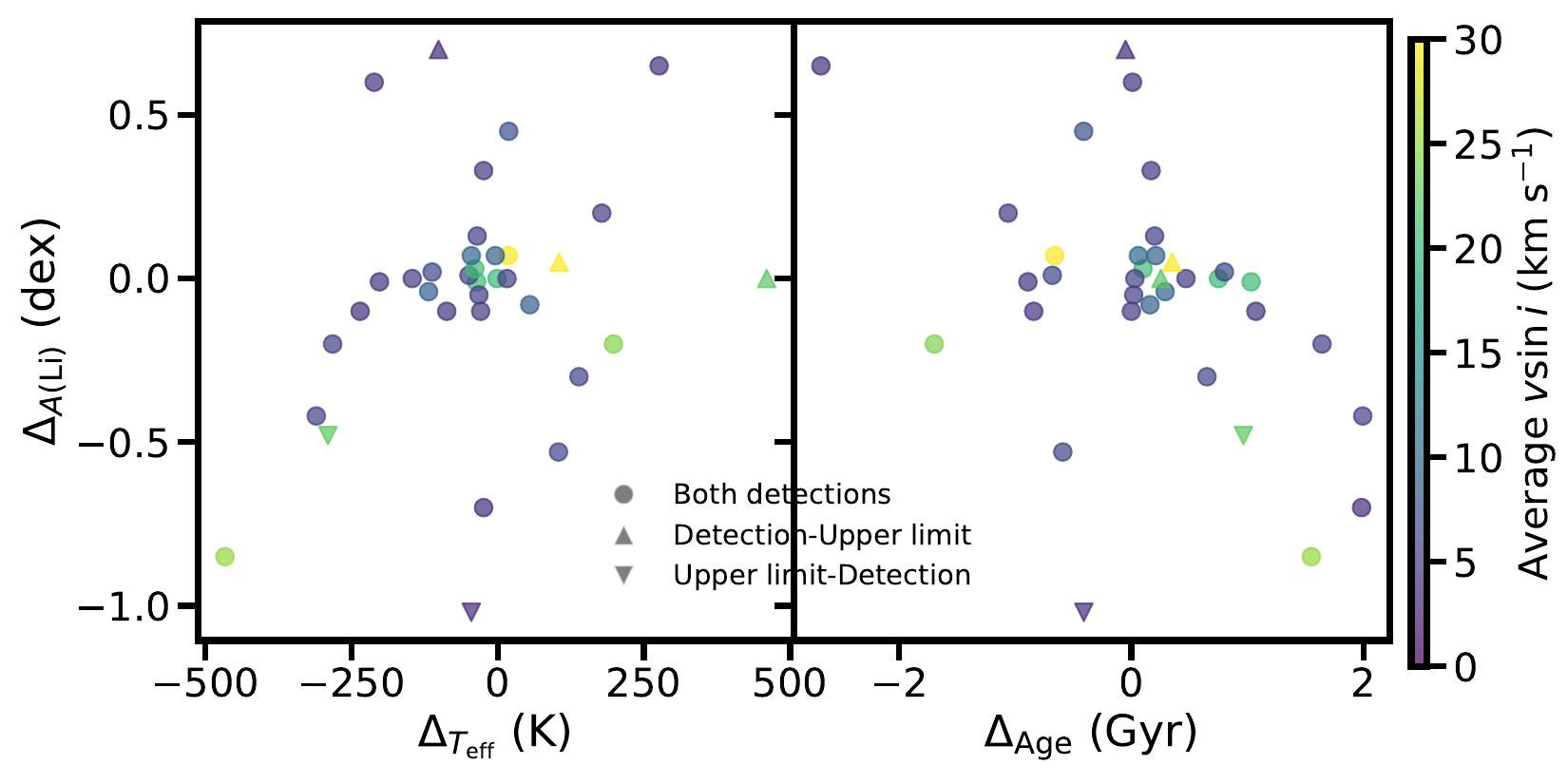}
	\caption{Differences in A(Li) ($\Delta$A(Li)) plotted against differences in 
$T_{\rm eff}$ ($\Delta\ T_{\rm eff}$) and age ($\Delta$age) for comoving pairs. Filled circles represent detection–detection pairs. Downward triangles indicate upper limit–detection pairs, where the true $\Delta$A(Li) is lower than shown; upward triangles represent detection–upper limit pairs, where the true value is higher than shown. Symbols are color-coded by the average {\it v} sin {\it i} of each pair.}
	\label{Fig4}
\end{figure*}

Although comoving pairs presumably share a common origin and age, some pairs exhibit differences in $T_{\rm eff}$ and age that could influence their A(Li) differences. Figure \ref{Fig4} presents $\Delta$A(Li) plotted against $\Delta\ T_{\rm eff}$ and $\Delta$age for the comoving pairs. We find no consistent correlation between either $\Delta\ T_{\rm eff}$ or $\Delta$age and $\Delta$A(Li), indicating the absence of a temperature- or age-dependent trend. This supports the conclusion that stellar rotation, rather than small differences in mass or $T_{\rm eff}$, is the primary factor driving the Li-Dip.

For pairs in which both stars have {\it v} sin {\it i} $<$ 12 km s$^{-1}$, the relative rotation rate does not consistently predict A(Li), as both components are slow rotators that have likely undergone varying spin-down histories, making it difficult to assess their respective A(Li) depletion rates. We note that the A(Li) -- {\it v} sin {\it i} relation shown here offers an instantaneous snapshot rather than a time-resolved sequence. While we measure present-day {\it v} sin {\it i}, the full rotational evolution of each star, including initial {\it v} sin {\it i} and spin-down rate, remains uncertain.
		
The rotation-Li depletion relationship in Li-Dip stars needs an important consideration. {\it v} sin {\it i} measurements are affected by unknown inclination angles. A star with {\it v} sin {\it i} = 30 km s$^{-1}$ could either be: (a) a fast rotator ($\sim$60 km s$^{-1}$) that has slowed down, or (b) a star maintaining its original rotation speed ($\sim$30 km s$^{-1}$) viewed equator-on. A few comoving pairs may distort the observed pattern due to the  sin {\it i} effect. However, $P_{\rm ROT}$ are not available for most of the stars in our sample.
		
\section{Discussion and Conclusions}
		
In this study, we present new high-precision stellar parameters and lithium abundances for 22 comoving pairs on the hotter side of the Li-Dip and slightly beyond. Combined with previously published C3PO data, our analysis includes 124 stars spanning the Li-Dip region ($T_{\rm eff}$ = 6735–6200 K) and extending slightly beyond (($T_{\rm eff}$=6000–7300 K). This sample contains 49 comoving pairs (98 stars) with both components falling in 6000-7300 K.

We examine the A(Li)–$T_{\rm eff}$ relationship across three distinct regions: (1) early F dwarfs ($T_{\rm eff}$ $>$ 6735 K), which show moderate Li depletion and rapid rotation; (2) the Li-Dip ($T_{\rm eff}$ = 6735–6200 K), characterized by severe Li depletion and large scatter; and (3) the Li Plateau ($T_{\rm eff}$ = 6200–6000 K), where A(Li) trends are more tight but still show evidence of age-dependent depletion.

We observe a correlation between A(Li) and stellar rotation for the three pairs highlighted with linked red lines in Figure~\ref{Fig3}. In these comoving pairs, where one component has {\it v} sin {\it i} $>$ 12 km s$^{-1}$ and the other has {\it v} sin {\it i} $<$ 12 km s$^{-1}$, the faster rotator consistently shows higher A(Li). This pattern suggests that rotational spin-down plays a role in Li depletion among mid-F dwarfs. Additionally, in pairs where both stars are slow rotators, we find lower A(Li) and larger pair-to-pair variation compared to the fast rotators, implying that differences in spin-down histories may contribute to the observed Li dispersion even within comoving systems.
		
In the hottest region ($T_{\rm eff}$ $\sim$ 7200 K), stars retain relatively high A(Li) until they evolve off the main sequence, at which point structural changes trigger spin-down and lithium depletion. At $T_{\rm eff}$ $\sim$ 6700 K, A(Li) declines from 2.5 to 1.7 dex, independent of metallicity. This supports that internal structural changes, rather than atmosphere parameters like $T_{\rm eff}$, log g, are responsible for Li depletion in this region. Since stars near $T_{\rm eff}$ $\sim$ 6700 K possess fully radiative envelopes, magnetic braking is unlikely to be the primary mechanism. Instead, the observed correlation between rotation and A(Li) suggests that internal processes, particularly the coupling between a slowing convective core and a radiative envelope, drive angular momentum transport and mixing from the interior to the surface.
		
In the central and cooler portions of the Li-Dip (6500–6200 K), stars develop increasing convective outer layers, enabling stronger magnetic braking and progressive Li depletion with age. However, the fact that the Li-Dip weakens toward cooler temperatures suggests additional processes are involved. This region marks the structural transition from stars with radiative envelopes and convective cores to those with convective envelopes and radiative cores, likely altering both spin-down efficiency and Li depletion rates. The changing internal structure may weaken angular momentum coupling between the stellar interior and envelope, moderating further Li depletion. Alternatively, stars in this temperature range may reach the main sequence already partially spun down due to extended pre-main-sequence evolution, resulting in reduced angular momentum loss and interior mixing during the main-sequence phase.
		
Our comoving pairs provide improved A(Li) detection limits compared to previous open cluster studies, offering valuable constraints on stellar models of angular momentum loss and internal mixing in F-type stars. The correlation between rotation and A(Li) across comoving pairs highlights that rotational history is as important as age and mass in driving Li depletion. Several stellar evolution models have already incorporated angular momentum transport and rotationally induced mixing \citep[e.g.,][]{2021A&A...646A..48D, 2011ApJS..192....3P}, and have successfully reproduced some observed trends. Future observations targeting more comoving pairs across a range of ages, particularly those ``twins" with minimal temperature differences, will help further constrain the formation and evolution of the Li-Dip and the internal mixing processes that drive it.
		
\section*{Acknowledgements}
		
We acknowledge Marc Pinsonneault and Vincent Smedile for helpful discussions that helped improve this paper. Q.S. is supported by the National Key R\&D Program of China No. 2024YFA1611801, the Science and Technology Commission of Shanghai Municipality under Grant No. 25ZR1402244, and the Startup Fund for Young Faculty at Shanghai Jiao Tong University. Y.S.T is supported by the National Science Foundation under Grant No. 2406729. This work has made use of data from the European Space Agency (ESA) mission Gaia (\url{https://www.cosmos.esa.int/gaia}), processed by the Gaia Data Processing and Analysis Consortium (DPAC, \url{https://www.cosmos.esa.int/web/gaia/dpac/consortium}). Funding for the DPAC has been provided by national institutions, in particular, the institutions participating in the Gaia Multilateral Agreement. This paper includes data gathered with the 6.5-m Magellan Telescopes located at Las Campanas Observatory, Chile, kindly supported by Carnegie Observatories. Some of the data presented herein were obtained at the W. M. Keck Observatory, which is operated as a scientific partnership among the California Institute of Technology, the University of California, and the National Aeronautics and Space Administration. The Observatory was made possible by the generous financial support of the W. M. Keck Foundation. Based on observations collected at the European Southern Observatory under ESO programme 108.22EC.001.
		
\bibliography{fast_Li}{}
\bibliographystyle{aasjournal}

\appendix 

Table \ref{TableA1} lists the stellar parameters for 22 pairs newly observed in this work. The columns include our designated ID, SIMBAD name, right ascension and declination (J2000), effective temperature ($T_{\rm eff}$), surface gravity (log {\it g}), metallicity ([Fe/H]), microturbulence ($V_t$), and their respective uncertainties. Also provided are the projected rotational velocity ({\it v} sin {\it i}) and uncertainty, age and uncertainty, as well as the signal-to-noise ratio, the 1D LTE lithium abundance (A(Li)) with its uncertainty, and the non-LTE correction to A(Li).

		\begin{longrotatetable}
			\begin{deluxetable*}{ccp{1.5cm}p{1.5cm}cccccccccccccccc}
				\renewcommand\thetable{A1}
				\tablecaption{Stellar Parameters for 22 New Comoving Pairs 	\label{TableA1}}
				\tabletypesize{\small}
				\tablewidth{1.0\textwidth}
				\setlength{\tabcolsep}{0.4pt}
				\decimalcolnumbers
				\renewcommand{\arraystretch}{1.4}
				\tablehead{
					ID$^a$	 &  Name$^a$	&    RA$^a$	 &    DEC$^a$	& $T_{\rm eff}^b$	& log {\it g}$^b$ &   [Fe/H]$^b$ &  $V_t^b$  & $\sigma_{\rm Teff}^b$ &	$\sigma_{\rm logg}^b$	& $\sigma_{\rm [Fe/H]}^b$	& $\sigma_{\rm Vt}^b$	& {\it v} sin {\it i}$^c$	& $\sigma_{\rm vsini}^c$ &	 age$^c$	& $\sigma_{age}^c$ & S/N$^d$	& A(Li)$^e$	 & $\sigma_{ALi}^e$ & NLTE$_{cor}^e$ \\
					&  	&    deg	    &   deg	& K	&  &   dex &  km s$^{-1}$  & K &		& dex	&  km s$^{-1}$	&  km s$^{-1}$	&  km s$^{-1}$ &	 Gyr & Gyr &  & dex	 & dex & dex
				}
				\startdata
				\hline
				star9-a	    &  CD-30 10115	& 191.8627	& -30.8984	& 5824	& 4.54 &  0.043	& 1.02 &	 39	         &   0.10	        &    0.03	        & 0.10	        &   3.66	    & 0.02	          &  3.29	& 1.87	         & 137	& 2.90	    &   0.04	     & -0.12    \\
				star10-b	&  HD 109987	& 189.8045	& -29.1724	& 5880	& 4.57 &  0.035	& 1.18 &	 47	         &   0.11	        &    0.04	        & 0.11	        &   3.44	    & 0.02	          &  2.99	& 1.72	         & 81	& 2.90	    &   0.05	     & -0.11    \\
				star15-a*	&  HD 115600	& 199.8318	& -59.4723	& 7243	& 4.32 & -0.118	& 2.18 &    306	         &   0.50	        &    0.15	        & 0.37	        &  105.37	    & 1.87	          &  0.02-0.025	& --	         & 281	& 3.15	    &   0.10	     & -0.05    \\
				star16-b*	&  HD 114082	& 197.3174	& -60.3083	& 6945	& 4.56 & -0.023	& 1.70 &    226	         &   0.28	        &    0.11	        & 0.31	        &  41.51	    & 0.55	          & 0.02-0.025	& --	         & 266	& 3.20	    &   0.12	     & -0.04    \\
				star25-a*	&  HD 122281	& 210.7300	& -60.3459	& 6504	& 4.39 &  0.050	& 1.60 &    122	         &   0.17	        &    0.06	        & 0.18	        &    4.0	    & 0.03	          &  0.03-0.05	& --	         & 234	& 3.03	    &   0.04	     & -0.05    \\
				star26-b*	&  HD 123058	& 211.8720	& -61.5626	& 7002	& 4.54 & -0.142	& 2.07 &    315	         &   0.42	        &    0.15	        & 0.43	        &   93.3	    & 1.74	          &  0.03-0.05	& -- & 318	& 3.23	    &   0.14	     & -0.03    \\
				star45-a	&  CD-75 836	& 236.0196	& -75.7062	& 6435	& 4.12 &  0.234	& 1.81 &	 47	         &   0.15	        &    0.05	        & 0.11	        &    4.1	    & 0.02	          &  1.74	& 0.54	         & 215	& $<$2.00	&   --	         &  --      \\
				star46-b	&  HD 139151	& 235.8768	& -75.7099	& 6407	& 4.12 &  0.294	& 1.73 &	 31	         &   0.10	        &    0.03	        & 0.07	        &    3.8	    & 0.04	          &  1.80	& 0.44	         & 205	& 2.25	    &   0.06	     & -0.01    \\
				star49-a*	&  HD 141813	& 237.9763	& -26.3676	& 6062	& 4.41 & -0.116	& 1.99 &    128	         &   0.19	        &    0.08	        & 0.24	        &   40.7	    & 0.42	          &  0.005-0.020 & -- & 303	& 3.10	    &   0.11	     & -0.09    \\
				star50-b*	&  HD 142505	& 238.9385	& -24.6819	& 6408	& 4.58 & -0.005	& 1.54 &    162	         &   0.18	        &    0.08	        & 0.25	        &   31.4	    & 0.26	          &  0.005-0.020 & -- & 308	& 3.17	    &   0.11	     & -0.06    \\
				star1-a*	&  HD 104638	& 180.7494	& -10.7519	& 6680	& 4.50 & -0.005	& 1.60 &    135	         &   0.17	        &    0.07	        & 0.20	        &   8.43	    & 0.09	          &  3.03	& 0.65	         & 351	& 2.20	    &   0.08	     & -0.01    \\
				star2-b*    &  HD 113103	& 195.4035	& -24.6657	& 7146	& 4.50 & -0.064	& 1.99 &    222	         &   0.32	        &    0.11	        & 0.30	        &  42.09	    & 0.52	          &  1.48	& 0.56	         & 425	& 3.05	    &   0.10	     & -0.02    \\
				star5-a*    &  HD 107145	& 184.9366	& -76.8014	& 6704	& 4.27 &  0.048	& 1.83 &	134	         &   0.20  	        &    0.07     	    & 0.17	        &   8.32	    & 0.09	          &  2.31	& 0.90	         & 337	& $<$1.70	&   --	         &  --      \\
				star6-b*    &  HD 121187	& 208.6593	& -34.5975	& 6994	& 4.49 &  0.102	& 1.92 &	177	         &   0.25     	    &    0.09	        & 0.24	        &  36.98	    & 0.36	          &  1.34	& 0.55	         & 301	& $<$2.30	&   --	         &  --      \\
				star7-a*    &  HD 107892	& 186.0088	& -19.5716	& 6831	& 4.47 &  0.055	& 1.71 &	187	         &   0.25	        &    0.09	        & 0.25	        &  40.14	    & 0.43	          &  1.70	& 0.67	         & 315	& $<$2.20	&   --	         &  --      \\
				star8-b*    &  HD 121852	& 209.7903	& -45.4687	& 6371	& 4.19 &  0.035	& 1.60 &     91	         &   0.13	        &    0.05	        & 0.11	        &   4.52	    & 0.03	          &  1.44	& 0.66	         & 288	& $<$2.00	&   --	         &  --      \\
				star11-a*	&  HD 114001	& 196.9120	& -14.1874	& 6731	& 4.56 &  0.105	& 1.58 &    157	         &   0.19	        &    0.08	        & 0.24	        &  35.31	    & 0.23	          &  1.14	& 0.49	         & 327	& 2.40	    &   --	         & -0.02    \\
				star12-b*	&  HD 125436	& 214.8502	& -02.0977	& 6533	& 4.51 &  0.108	& 1.54 &    121	         &   0.15	        &    0.06	        & 0.18	        &  13.29	    & 0.09	          &  2.83	& 0.98	         & 293	& 2.60	    &   0.09	     & -0.02    \\
				star19-a*	&  HD 117620	& 203.1770	& -55.8274	& 7133	& 4.04 & -0.286	& 2.79 &    396	         &   0.58	        &    0.20	        & 0.45	        &  91.22	    & 1.44	          &  0.020-0.025	& -- & 286	& 3.20	    &   0.18	     & -0.04    \\
				star20-b*	&  HD 113556	& 196.3859	& -58.5355	& 7266	& 4.71 &  0.052	& 2.01 &    233	         &   0.33	        &    0.11	        & 0.34	        &   61.1	    & 0.99	          &  0.020-0.025	& --  & 290	& 3.02	    &   0.14	     & -0.04    \\
				star23-a*	&  HD 119718	& 206.6477	& -62.0691	& 6857	& 3.60 & -0.408	& 2.86 &    389	         &   0.57	        &    0.21	        & 0.44	        &   70.3	    & 1.61	          &  0.020-0.025 & -- & 297	& 3.25	    &   0.16	     & -0.04    \\
				star24-b*	&  HD 117214	& 202.5375	& -58.4843	& 6627	& 4.45 &  0.033	& 1.46 &    174	         &   0.22	        &    0.08	        & 0.48	        &   42.4	    & 0.25	          &  2.70	& 0.92	         & 293	& 3.20	    &   0.14	     & -0.05    \\
				star51-a	&  HD 147814	& 246.9905	& -62.5986	& 6546	& 3.99 & -0.187	& 1.98 &	 51	         &   0.19	        &    0.04	        & 0.15	        &    4.4	    & 0.03	          &  2.24	& 0.51	         & 257	& $<$2.05	&   --	         &  --      \\
				star52-b	&  HD 160416	& 265.5718	& -49.2701	& 6633	& 4.44 & -0.030	& 1.85 &	 47	         &   0.15	        &    0.04	        & 0.13	        &    4.8	    & 0.03	          &  1.17	& 0.60	         & 204	& $<$2.10	&   --	         &  --      \\
				star53-a*	&  HD 148484	& 247.7238	& -50.9706	& 7152	& 4.06 & -0.029	& 2.46 &    206	         &   0.30	        &    0.10	        & 0.24	        &   36.6	    & 0.46	          &  1.13	& 0.53	         & 317	& 3.25	    &   0.14	     & -0.05    \\
				star54-b*	&  HD 170691	& 279.1190	& -69.0815	& 7047	& 4.05 & -0.122	& 2.50 &    275	         &   0.42	        &    0.14	        & 0.32	        &   60.3	    & 0.93	          &  0.77	& 0.48	         & 195	& $<$3.20	&   --	         &  --      \\
				star59-a	&  HD 171413	& 278.8355	& -16.1411	& 6499	& 4.02 & -0.095	& 1.97 &	 53	         &   0.17	        &    0.05	        & 0.14	        &    7.3	    & 0.05	          &  2.24	& 0.51	         & 327	& $<$2.05	&   --	         &  --      \\
				star60-b	&  HD 166761	& 273.0993	& -7.2973	& 6288	& 4.19 & -0.046	& 1.52 &	 30	         &   0.09	        &    0.02	        & 0.07	        &    4.2	    & 0.04	          &  2.93	& 0.61	         & 332	& $<$1.70	&   --	         &  --      \\
				star71-a*	&  HD 191089	& 302.2718	& -26.2240	& 6669	& 4.60 & -0.009	& 1.62 &	215	         &   0.24	        &    0.10	        & 0.30	        &   40.3	    & 0.58	          &  0.020-0.025 & --	         & 151	& 3.25	    &   0.09	     & -0.06    \\
				star72-b*	&  HD 164249	& 270.7644	& -51.6488	& 6748	& 4.51 &  0.024	& 1.66 &    148	         &   0.19	        &    0.07	        & 0.20	        &   16.2	    & 0.16	          & 0.020-0.025	& --  & 383	& 3.17	    &   0.05	     & -0.03    \\
				star79-a	&  HD 206395	& 325.7606	& -43.4961	& 6293	& 4.40 &  0.235	& 1.39 &	 34	         &   0.09	        &    0.03	        & 0.08	        &    3.5	    & 0.03	          &  1.24	& 0.65	         & 268	& 2.85	    &   0.04	     & -0.05    \\
				star80-b	&  HD 152260	& 255.2285	& -76.2187	& 6394	& 4.33 &  0.242	& 1.56 &	 37	         &   0.10	        &    0.03	        & 0.08	        &    3.8	    & 0.04	          &  1.29	& 0.60	         & 183	& $<$2.15	&   --	         &  --      \\
				star99-gb	&  * 18 Sco	    & 243.9060	& -8.3724	& 5818	& 4.43 &  0.039	& 0.78 &	 31	         &   0.08	        &    0.03	        & 0.08	        &    5.6	    & 0.02	          &  4.33	& 1.89	         & 220	& 1.60	    &   0.06	     & -0.06    \\
				star100-gb	&  * bet Vir	& 177.6793	& 1.7624	& 6270	& 4.29 &  0.183	& 1.30 &	 28	         &   0.08	        &    0.02	        & 0.06	        &    4.4	    & 0.03	          &  1.87	& 0.69	         & 161	& 2.13	    &   0.04	     & -0.03    \\
				star101-gb	&  HD 140283	& 235.7565	& -10.9339	& 5750	& 3.70 & -2.500	& 1.40 &	100	         &   0.10	        &    0.20	        & 0.10	        &    8.0	    & 0.29	          &  14.27	& 0.38	         & 357	& 2.28	    &   0.20	     & 0.00     \\
				star102-gb	&  * mu. Ara	& 266.0363	& -51.8344	& 5831	& 4.36 &  0.293	& 0.89 &	 33	         &   0.09	        &    0.03	        & 0.10	        &    5.0	    & 0.02	          &  3.25	& 1.25	         & 140	& $<$1.60	&   --	         &  --      \\
				star223-a	&  HD 145920	& 243.6960	& -33.2572	& 5866	& 4.37 &  0.235	& 0.90 &	 31	         &   0.08	        &    0.03	        & 0.09	        &    5.4	    & 0.02	          &  3.15	& 1.23	         & 130	& 2.26	    &   0.04	     & -0.06    \\
				star224-b	&  BD-02 3883	& 220.2784	& -3.4064	& 6404	& 4.55 &  0.172	& 1.50 &	 36	         &   0.11	        &    0.03	        & 0.09	        &    3.7	    & 0.03	          &  0.91	& 0.56	         & 129	& 3.01	    &   0.04	     & -0.06    \\
				star243-a	&  HD 183961	& 293.0579	& -5.1841	& 6382	& 4.29 & -0.011	& 1.52 &	 37	         &   0.11	        &    0.03	        & 0.08	        &    4.3	    & 0.04	          &  1.97	& 0.76	         & 198	& $<$2.20	&   --	         &  --      \\
				star244-b	&  HD 189924	& 300.6884	& -5.5102	& 6427	& 4.06 & -0.011	& 1.75 &	 36	         &   0.13	        &    0.03	        & 0.08	        &    3.8	    & 0.03	          &  2.38	& 0.46	         & 228	& 3.22	    &   0.03	     & -0.06    \\
				star4-b	    &  HD 109172	& 188.2567	& -41.2455	& 6437	& 4.38 & -0.165	& 1.59 &	 45	         &   0.15	        &    0.04	        & 0.11	        &    3.8	    & 0.03	          &  1.85	& 0.86	         & 145	& 2.67	    &   0.05	     & -0.04    \\
				star29-a	&  HD 126766	& 216.9378	& -13.3584	& 6419	& 4.40 & -0.138	& 2.45 &	129	         &   0.37	        &    0.10	        & 0.27	        &   71.2	    & 1.61	          &  2.51	& 1.10	         & 472	& 2.60	    &   0.05         & -0.04    \\
				star76-b	&  HD 152322	& 254.2761	& -64.1885	& 5975	& 4.44 & -0.038	& 0.97 &	 31	         &   0.08	        &    0.02	        & 0.06	        &    5.7	    & 0.02	          &  3.12	& 1.49	         & 223	& $<$1.70	&   --	         &  --      \\
				star315-a & TYC 1098-2147-1	& 312.7954	& 12.3671	& 6239	& 4.56 &  0.026	& 1.33 &	 43	         &   0.12	        &    0.04	        & 0.10	        &    6.3	    & 0.03	          &  1.49	& 0.85	         & 297	& 3.05	    &   0.04	     & -0.09    \\
				\hline
				\enddata\tablecomments{a. Our designated pair ID, Simbad name of the star, RA and DEC in degrees (J2000). An asterisk (*) following the pair ID indicates stellar parameters derived using the FASMA synthesis code (\citealt{2018MNRAS.473.5066T}).
					b. Stellar atmosphere parameters, including $T_{\rm eff}$, log {\it g}, $V_t$, and their associated uncertainties.
					c. Rotational velocity and associated uncertainty of the star; age and associated uncertainty of the star.
					d. Signal-to-noise ratio (S/N) near the Li I 6707.8 \AA doublet.
					e. The 1D-LTE A(Li), if the number starts with a ``$<$", it means an upper limit A(Li), else it is a detection. If an detection, the uncertainty and Non-LTE corrections associated with A(Li).
				}
			\end{deluxetable*}
		\end{longrotatetable}
	
	\end{document}